\documentstyle[12pt]{article}
\begin{document}
\def\ds{\displaystyle}
\def\beq{\begin{equation}}
\def\eeq{\end{equation}}
\def\bea{\begin{eqnarray}}
\def\eea{\end{eqnarray}}
\def\ve{\vert}
\def\vel{\left|}
\def\ver{\right|}
\def\nnb{\nonumber}
\def\ga{\left(}
\def\dr{\right)}
\def\aga{\left\{}
\def\adr{\right\}}
\def\lla{\left<}
\def\rra{\right>}
\def\rar{\rightarrow}
\def\nnb{\nonumber}
\def\la{\langle}
\def\ra{\rangle}
\def\ba{\begin{array}}
\def\ea{\end{array}}
\def\tr{\mbox{Tr}}
\def\ssp{{\Sigma^{*+}}}
\def\sso{{\Sigma^{*0}}}
\def\ssm{{\Sigma^{*-}}}
\def\xis0{{\Xi^{*0}}}
\def\xism{{\Xi^{*-}}}
\def\qs{\la \bar s s \ra}
\def\qu{\la \bar u u \ra}
\def\qd{\la \bar d d \ra}
\def\qq{\la \bar q q \ra}
\def\gGgG{\la g^2 G^2 \ra}
\def\q{\gamma_5 \not\!q}
\def\x{\gamma_5 \not\!x}
\def\g5{\gamma_5}
\def\sb{S_Q^{cf}}
\def\sd{S_d^{be}}
\def\su{S_u^{ad}}
\def\ss{S_s^{??}}
\def\sbp{{S}_Q^{'cf}}
\def\sdp{{S}_d^{'be}}
\def\sup{{S}_u^{'ad}}
\def\ssp{{S}_s^{'??}}
\def\sig{\sigma_{\mu \nu} \gamma_5 p^\mu q^\nu}
\def\fo{f_0(\frac{s_0}{M^2})}
\def\ffi{f_1(\frac{s_0}{M^2})}
\def\fii{f_2(\frac{s_0}{M^2})}
\def\O{{\cal O}}
\def\sl{{\Sigma^0 \Lambda}}
\def\es{\!\!\! &=& \!\!\!}
\def\ar{&+& \!\!\!}
\def\ek{&-& \!\!\!}
\def\cp{&\times& \!\!\!}
\def\se{\!\!\! &\simeq& \!\!\!}
\title{
         {\Large
                 {\bf
The magnetic moments of $\Lambda_b$ and $\Lambda_c$ baryons
in light cone QCD sum rules
                 }
         }
      }

\author{\vspace{1cm}\\
{\small T. M. Aliev \thanks
{e-mail: taliev@metu.edu.tr}\,\,,
A. \"{O}zpineci \thanks
{e-mail: altugoz@metu.edu.tr}\,\,,
M. Savc{\i} \thanks
{e-mail: savci@metu.edu.tr}} \\
{\small Physics Department, Middle East Technical University} \\
{\small 06531 Ankara, Turkey} }
\date{}

\begin{titlepage}
\maketitle
\thispagestyle{empty}

\begin{abstract}
Using the most general form of the interpolating currents of heavy baryons,
the magnetic moments of heavy baryons $\Lambda_Q~(Q=b,c)$ are calculated in
framework of the light cone QCD sum rules. 
A comparison of our results on magnetic moments with the existing 
theoretical results calculated in various other frameworks are presented.
\end{abstract}

~~~PACS number(s): 11.55.Hx, 13.40.Em, 14.20.Lq, 14.20.Mr
\end{titlepage}

\section{Introduction}

QCD sum rules \cite{R1} are very successful in determination of the masses
and coupling constants of low--lying mesons and baryons. In this method a
deep connection between hadron properties and QCD vacuum structure is
established via few condensates. This approach is adopted and extended to
many works (see for example \cite{R2} and the references therein). One of
the important static characteristics of baryons is their magnetic
moments. The magnetic moment of nucleon within QCD sum rules is obtained in
\cite{R3,R4} using external field method. In \cite{R5} use is made of the
QCD sum rules method in the presence of external electromagnetic field
with field strength tensor 
${\cal F}_{\mu\nu}$ to calculate the magnetic moments of the baryons
$\Sigma_c,~\Lambda_c$ containing heavy quarks. 

The goal of the present work is to calculate the magnetic moments of
$\Lambda_b$ and $\Lambda_c$ in framework of an alternative approach to the
traditional QCD sum rules, i.e., light cone QCD sum rules (LCQSR) (more 
about this method and its applications can be found in \cite{R6,R7} and
references therein).
Magnetic moments of the nucleons and decoupled baryons have been studied in
LCQSR in \cite{R8} and \cite{R9,R10}, respectively.  

The paper is organized as follows. In section 2 the LCQSR for
$\Lambda_Q~(Q=b,c)$ magnetic moment is derived. In section 3 we present
numerical results.

\section{LCQSR for $\Lambda_Q$ magnetic moment}

Our starting point for determination of the $\Lambda_Q$ magnetic moment is
to consider the two--point correlator function 

\bea
\label{pi1}
\Pi = i \int d^4 x \, e^{i p x} \lla 0 \ve T\{\eta_{\Lambda_Q}(x)
\bar \eta_{\Lambda_Q}(0) \} \ve 0 \rra_{ {\cal F}_{\alpha\beta}}~,
\eea
where ${\cal F}_{\alpha\beta}$ is the external electromagnetic field,
$\eta_{\Lambda_Q}$ is the interpolating current with $\Lambda_Q$ quantum
numbers. It is well known that there is a continuum of choices for the
baryon interpolating currents. The general form of $\Lambda_Q$ currents can
be written as \cite{R11}
\bea
\label{eta}
\eta_{\Lambda_Q} \es 2 \ga \eta_{\Lambda_1} + t \eta_{\Lambda_2}
\dr~, \nnb \\
\eea
where $t$ is an arbitrary parameter and
\bea
\label{eta1}
\eta_{\Lambda_1} \es \frac{1}{\sqrt{6}} \epsilon_{abc}
\left[ 2 (u_a^T C d_b) \gamma_5 Q_c + (u_a^T C Q_b) \gamma_5 d_c - (d_a^T
C Q_b) \gamma_5 u_c \right]~, \\
\label{eta2}
\eta_{\Lambda_2} \es \frac{1}{\sqrt{6}} \epsilon_{abc}
\left[ 2 (u_a^T C \gamma_5 d_b) Q_c + (u_a^T C \gamma_5 Q_b) d_c -
(d_a^T C \gamma_5 Q_b) u_c \right] ~,
\eea
where $a$, $b$, and $c$ are color indices. Ioffe current corresponds to the
choice $t=-1$.

Let us consider phenomenological part of the correlator (\ref{eta}).
Saturating the correlator with the complete set of hadron states having the
same quantum numbers with $\Lambda_Q$ baryon, we get
\bea
\label{pi2}
\Pi = \sum_i \frac{\lla 0 \ve \eta_{\Lambda_Q} \ve B_i (p_1) \rra}{p_1^2 -
M^2} \lla B_i(p_1) \ve B_i (p_2) \rra_{ {\cal F}_{\alpha\beta}}
\frac{\lla B_i(p_2) \ve \bar \eta_{\Lambda_Q} \ve 0 \rra}{p_2^2 - M^2}~,
\eea
where $p_2 = p_1 + q$, $q$ is the photon momentum and $B_i$ is the complete set
of corresponding baryons having the same quantum numbers as $B$ with masses
$M$.

The interpolating current couples to the baryon states with
amplitudes $\lambda$ defined as
\bea
\label{amp1}
\lla 0 \ve \eta_{\Lambda_Q} \ve \Lambda_Q \rra = \lambda_{\Lambda_Q}
u_{\Lambda_Q} (p)~.
\eea
It follows from Eq. (\ref{pi2}) that in order to calculate the 
phenomenological part of the correlator, an expression for the matrix
element $\lla B (p_1) \ve B (p_2) \rra_{ {\cal F}_{\alpha\beta}}$ is needed.
This matrix element can be parametrized as follows:
\bea
\label{amp2}
\lla B(p_1) \ve B(p_2) \rra_{ {\cal F}_{\alpha\beta}} \es \bar u (p_1)
\left[ f_1 \gamma_\mu + i \, \frac{\sigma_{\mu\alpha} q^\alpha}{2
m_{\Lambda_Q}}
f_2 \right] u (p_2) \varepsilon^\mu~, \nnb \\
\es \bar u (p_1) \left[ (f_1 + f_2) \gamma_\mu + \frac{(p_1 + p_2)_\mu}
{2 m_{\Lambda_Q}} f_2 \right] u (p_2) \varepsilon^\mu~,
\eea
where $f_1$ and $f_2$ are the form factors, which are functions of $q^2 = (p_2 -
p_1)^2$ and $\varepsilon^\mu$ is the polarization four vector of the photon.
In the present case, in order to calculate magnetic
moment of $\Lambda_Q$, the values of the form factors at $q^2=0$ are needed.
Using Eqs. (\ref{pi2})--(\ref{amp2}), for the phenomenological
part of the LCQSR we get:
\bea
\label{pi3}
\Pi = - \lambda_{\Lambda_Q}^2 \varepsilon^\mu
\frac{\not\!p_1 + m_{\Lambda_Q}}{p_1^2 - m_{\Lambda_Q}^2}
\left[ (f_1 + f_2) \gamma_\mu + \frac{(p_1 +
p_2)_\mu}{2 m_{\Lambda_Q}}
f_2 \right] \frac{\not\!p_2 + m_{\Lambda_Q}}{p_2^2 -
m_{\Lambda_Q}^2}~.
\eea
Among a number of different structures present in Eq. (\ref{pi3}), we choose
$\not\!p_1\!\!\not\!\varepsilon\!\!\not\!p_2$ which contains the magnetic moment
form factor $f_1+f_2$. When calculated at $q^2=0$, this structure gives
the magnetic moment of $\Lambda_Q$ baryon in units of $e\hbar/2m_{\Lambda_Q}$.
Isolating the phenomenological part of the correlator from this structure
which describes the magnetic moment of the $\Lambda_Q$ baryon, we get
\bea
\label{pi4} 
\Pi = - \lambda_{\Lambda_Q}^2 \frac{1}{p_1^2 -
m_{\Lambda_Q}^2}
\mu_{\Lambda_Q} \frac{1}{p_2^2 - m_{\Lambda_Q}^2}~,
\eea
where $\mu_{\Lambda_Q} = (f_1 + f_2) \ve_{q^2=0}$.

According to the QCD sum rules philosophy in order to construct sum 
rules we need to calculate the theoretical part of the correlator 
function $\Pi$. Calculating correlator (\ref{pi1}) in QCD we get


\bea
\label{cor}
\lefteqn{
\Pi(p_1^2,p_2^2) = - \frac{2}{3}\,\epsilon^{abc}
\epsilon^{def} 
\int d^4x \, e^{i p x} } \nnb \\ 
\cp  \lla 0 \vel  \Big\{ 
4 \g5 \sb \g5 \tr \sd \sup + 
4 t \g5 \sb \tr \sd \g5 \sup   \right. \right. 
\nnb \\ 
\ar 4 t \sb \g5 \tr \sd \sup \g5 + 
4 t^2 \sb \tr \sd \g5 \sup \g5 
\nnb \\ 
\ar
2 \g5 \sb \sup \sd \g5 +
2 t \g5 \sb \g5 \sup \sd
\nnb \\ 
\ar
2 t \sb \sup \g5 \sd \g5 +
2 t^2 \sb \g5 \sup \g5 \sd
\nnb \\ 
\ar
2 \g5 \sb \sdp \su \g5 +
2 t \g5 \sb \g5 \sdp \su
\nnb \\ 
\ar
2 t \sb \sdp \g5 \su \g5 +
2 t^2 \sb \g5 \sdp \g5 \su
\nnb \\ 
\ar
2 \g5 \sd \sup \sb \g5 +
2 t \g5 \sd \g5 \sup \sb
\nnb \\ 
&&+
2 t \sd \sup \g5 \sb \g5 +
2 t^2 \sd \g5 \sup \g5 \sb
\nnb \\ 
\ar
\g5 \sd \g5 \tr \sb \sup  +
t \g5 \sd \tr \sb \g5 \sup
\nnb \\ 
&&+
t \sd \g5 \tr \sb \sup \g5 +
t^2 \sd \tr \sb \g5 \sup \g5
\nnb \\ 
\ek
\g5 \sd \sbp \su \g5 -
t \g5 \sd \g5 \sbp \su
\nnb \\ 
\ek
t \sd \sbp \g5 \su \g5 -
t^2 \sd \g5 \sbp \g5 \su    
\nnb \\
\ar
2 \g5 \su \sdp \sb \g5 +
2 t \g5 \su \g5 \sdp \sb    
\nnb \\
\ar
2 t \su \sdp \g5 \sb \g5 +   
2 t^2 \su \g5 \sdp \g5 \sb    
\nnb \\
\ek
\g5 \su \sbp \sd \g5 -
t \g5 \su \g5 \sbp \sd    
\nnb \\
\ek
t \su \sbp \g5 \sd \g5 -   
t^2 \su \g5 \sbp \g5 \sd    
\nnb \\
\ar
\g5 \su \g5 \tr \sb \sdp +
t \g5 \su \tr \sb \g5 \sdp    
\nnb \\
\ar
t \su \g5  \tr \sb \sdp \g5 +   
t^2 \su \tr \sb \g5 \sdp \g5 
\left. \left. \Big\} \ver 0  \rra_{ {\cal F}_{\alpha\beta} }~,
\eea
where $S^{'} = C S^T C$, with $C$ and $T$ are being the charge conjugation and
transpose of the operator, respectively.

The perturbative contribution (i.e., photon is
radiated from the freely propagating quarks) can easily be obtained by 
making the following substitution in one of the propagators in Eq. (\ref{cor})
\bea
\label{sss}
S^{ab}_{\alpha \beta} \rightarrow 2 \left( \int dy \, {\cal F}^{\mu \nu}
y_\nu S^{free} (x-y) \gamma_\mu
S^{free}(y) \right)^{ab}_{\alpha \beta}~,
\eea
where the Fock--Schwinger gauge $x_\mu A^\mu(x) = 0$ and $S^{free}$ is the
free quark operator. In $x$--representation the propagator of the free
massive quark is 
\bea
\label{fmq}
S_Q^{free} = \frac{m_Q^2}{4 \pi^2} \frac{K_1 \ga m_Q \sqrt{-x^2}
\dr}{\sqrt{-x^2}} - i \, \frac{m_Q^2 \not\!x}{4 \pi^2 x^2} K_2 
\ga m_Q \sqrt{-x^2} \dr~,
\eea
where $m_Q$ is the heavy quark mass and $K_i$ are the Bessel functions.  
Using the expansions for the Bessel functions 
\bea
K_1(x) &\sim& \frac{1}{x} + {\cal O} (x)~,\nnb \\
K_2(x) &\sim& \frac{2}{x^2} - \frac{1}{2} + {\cal O} (x^2)~, \nnb
\eea
and formally setting $m_Q \rar 0$, one can obtain the well known expression
of the free propagator for massless quark in
$x$ representation:
\bea
\label{frp}
S_q^{free} = \frac{i \not\!x}{2 \pi^2 x^4}~.
\eea
The nonperturbative contributions can be obtained from Eq. (\ref{cor}) by
replacing one of the propagators of light quarks with
\bea
\label{rep}
S^{ab}_{\alpha \beta} \rar - \frac{1}{4} \bar q^a A_j q^b \ga A_j \dr_{\alpha
\beta}~,
\eea
where $A_j = \Big\{ 1,~\gamma_5,~\gamma_\alpha,~i\gamma_5 \gamma_\alpha,
~\sigma_{\alpha \beta} /\sqrt{2}\Big\}$ and sum over $A_j$ is implied.	

The complete light cone expansion of the light quark propagator in external
field is calculated in \cite{R12}. It receives
contributions from the nonlocal operators 
$\bar q G q$, $\bar q G G q$, $\bar q q \bar q q$,
where G is the gluon field strength
tensor.  In this work we consider operators with only one gluon
field and neglect terms with two gluons $\bar q GG q$, and four
quarks $\bar q q \bar q q$, and this action
can be justified on the basis of an expansion in
conformal spin \cite{R13}.  In this approximation heavy and massless 
light quark propagators read
\bea
\label{hvy}
\lla 0 \vel T\Big\{\bar Q (x) Q(0)  \Big\} \ver 0 \rra \es
i S_Q^{free} (x) - i g_s \int \frac{d^4 k}{(2\pi)^4} \, e^{-ikx} \int_0^1
dv \Bigg[\frac{\not\!k + m_Q}{( m_Q^2-k^2)^2} \, G^{\mu\nu}(vx)
\sigma_{\mu\nu} \nnb \\
\ar \frac{1}{m_Q^2-k^2}\, v x_\mu G^{\mu\nu} \gamma_\nu \Bigg]~, \\ \nnb \\
\nnb \\
\lla 0 \vel T\Big\{\bar q(x) q(0)  \Big\} \ver 0 \rra \es
i \, \frac{\not\!x}{2 \pi^2 x^4} - \frac{\lla \bar q q \rra}{12} \ga 1 +
\frac{x^2 m_0^2}{16} \dr \nnb \\
\ek i \, g_s \int dv \, \Bigg[ \frac{\not\!x}{16 \pi^2 x^2}\, G^{\mu\nu}(vx)
\sigma_{\mu\nu} - \frac{i}{4  \pi^2 x^2}\, v x_\mu G^{\mu\nu} \gamma_\nu
\label{lgh}
\Bigg]~,
\eea
where $m_0$ is defined from the relation 
\bea
\lla \bar q i g_s G_{\mu\nu} \sigma^{\mu\nu} q \rra = m_0^2 \qq~, \nnb
\eea
and the operators in local part 
with dimension $d>5$ are
not taken into consideration since their contribution is negligible. 

It follows from Eqs. (\ref{cor})--(\ref{lgh}) that, in order to calculate the 
theoretical part of the correlator function the matrix elements of
non--local operators between photon and the vacuum state are needed. 
Up to twist--4, the photon wave functions are defined in the following way
\cite{R13}--\cite{R15}:
\bea
\label{pwf}
\la \gamma (q) \ve \bar q \gamma_\alpha \gamma_5 q \ve 0 \ra \es \frac{f}{4}
e_q \epsilon_{\alpha \beta \rho \sigma} \varepsilon^\beta
q^\rho x^\sigma \int_0^1 du \, e^{i u qx} \psi(u)~, \nnb \\
\la \gamma (q) \ve \bar q \sigma_{\alpha \beta} q \ve 0 \ra \es
i e_q \qq \int_0^1 du \, e^{i u q x} \Bigg\{ (\varepsilon_\alpha q_\beta -
\varepsilon_\beta q_\alpha) \Big[ \chi \phi(u) + x^2 \Big(g_1(u) - g_2(u)\Big)
\Big]  \nnb \\
\ar \Big[ qx (\varepsilon_\alpha x_\beta - \varepsilon_\beta x_\alpha) +
\varepsilon x (x_\alpha q_\beta - x_\beta q_\alpha) \Big] g_2
(u) \Bigg\}~,
\eea
where $\chi$ is the magnetic susceptibility of the quark condensate, $e_q$
is the quark charge, the functions $\phi(u)$ and $\psi(u)$
are the leading twist--2 photon wave functions, while $g_1(u)$ and $g_2(u)$
are the twist--4 functions.  Note that twist--3 photon wave functions are
all neglected in further calculations since their contribution changes the
results about $5\%$. 

The theoretical part of the correlator can be obtained by substituting
photon wave functions and expressions of light and heavy quark propagators
into Eq. (\ref{cor}). The sum rules is obtained by equating the
phenomenological and theoretical parts of the correlator. In order to
suppress the contributions of higher states and of continuum (for more
details, see \cite{R9,R10,R16,R17}) we perform double Borel transformations
of the variables $p_1^2=p^2$ and $p_2^2=(p+q)^2$ on both sides of the
correlator. It should be mentioned here that the Borel transformations for
$K_\nu(x)$ functions, which appear in the propagator of a massive quark,
were calculated in \cite{R18}. After lengthy calculations we get the 
following sum rules for the $\Lambda_Q$ magnetic moment:
\bea
\lefteqn{
\label{sm1}
\mu_{\Lambda_Q} \lambda_{\Lambda_Q}^2\, e^{-m_{\Lambda_Q}^2/M^2}
 =  \frac{1}{32 \pi^4} M^6 
\Big[(1-t)^2 (e_u+e_d) \Psi(2,-1,m_Q^2/M^2) } \nnb \\
\ar (13 + 10 t +13 t^2) e_Q \Psi(3,0,m_Q^2/M^2) \Big] \nnb \\
\ar \frac{1}{48 \pi^2} (1-t)^2 (e_u+e_d) M^4 
f \psi(u_0) \Psi(1,-1,m_Q^2/M^2) \nnb \\
\ar \frac{m_Q}{24 \pi^2} (-5+4 t+t^2) M^4 \qq (e_u+e_d) \chi \varphi(u_0)
\Psi(2,0,m_Q^2/M^2) \nnb \\
\ek \frac{m_Q}{36} (-5+4 t+t^2) \qq (e_u+e_d) f \psi(u_0) F_5 (m_Q^2/M^2)\nnb \\
\ar \frac{m_Q}{144 M^2} m_0^2 \qq (e_u+e_d) f \psi(u_0)
\Big\{ (-5+4 t+t^2) \Big[F_4 (m_Q^2/M^2)-F_5 (m_Q^2/M^2)\Big] \nnb \\
&&\,+3 (-1+t^2)F_5 (m_Q^2/M^2)\Big\} \nnb \\
\ar \frac{1}{18 \pi^2} \qq (e_u+e_d)\Big[ g_1(u_0)-g_2(u_0)\Big] 
\Big[4 \pi^2 (1-2 t + t^2) \qq F_4 (m_Q^2/M^2) \nnb \\
\ek 3 (-5 + 4 t+ t^2) m_Q M^2 \Psi(1,0,m_Q^2/M^2) \Big]\nnb \\
\ek \frac{m_0^2}{36 M^2} \qq^2 (e_u+e_d) \Big[ g_1(u_0)-g_2(u_0)\Big]
\Big\{2 (1-2 t+t^2) F_1 (m_Q^2/M^2)  \\
\ar (-5+2 t +3 t^2) \Big[F_4 (m_Q^2/M^2)-F_5 (m_Q^2/M^2)\Big]\Big\} \nnb \\
\ar \frac{1}{144} \qq^2 (e_u+e_d)\chi \varphi(u_0)
\Big[-8 (1-2 t+t^2) M^2 F_5 (m_Q^2/M^2)\nnb \\
\ar 2 m_0^2(1-2 t+t^2) F_4 (m_Q^2/M^2)
+  m_0^2 (-5 + 2 t + 3 t^2)F_5 (m_Q^2/M^2) \Big] \nnb \\
\ar \frac{m_0^2}{36 M^2} e_Q \qq^2 (-11-2 t+13 t^2)
\Big[F_4 (m_Q^2/M^2)-F_5 (m_Q^2/M^2)\Big] \nnb \\
\ar \frac{m_Q}{24 \pi^2} M^2 \qq \Big[(e_u+e_d) (-5+4 t+t^2)
\Psi(1,0,m_Q^2/M^2) \nnb \\
\ar 2 e_Q(-1-4 t +5 t^2) \Psi(2,1,m_Q^2/M^2)\Big] \nnb \\
\ar \frac{1}{18} e_Q \qq^2 (11+2 t-13 t^2) F_5 (m_Q^2/M^2)\nnb \\
\ek \frac{m_Q}{96 \pi^2} m_0^2 \qq (e_u+e_d) \Big[(-5+4 t+t^2) 
F_5 (m_Q^2/M^2)+ 
3 (-1+t^2) \Psi(1,1,m_Q^2/M^2)\Big] \nnb \\
\ar \frac{m_Q}{288 \pi^2} e_Q m_0^2 \qq (3+12 t-16 t^2)\Psi(1,1,m_Q^2/M^2)~.\nnb 
\eea
Here the functions $\Psi(\alpha,\beta,x)$ and $F_i(x)$ are
defined as 
\bea
\label{psi}
\Psi (\alpha,\beta,x) \es \frac{1}{\Gamma(\alpha)} 
\int_1^\infty dt \, e^{-t x} \, t^{\beta-\alpha-1}
(t-1)^{\alpha-1}~,~~~~~~(\alpha > 0)~, \nnb \\ 
F_1 (x) \es \ga x^2 - 2 x \dr e^{-x}~, \nnb \\
F_2 (x) \es  \ga x^2 - 4 x + 2 \dr e^{-x}~, \nnb \\
F_3 (x) \es \ga x^2 - 6 x + 6 \dr e^{-x}~, \nnb \\
F_4 (x) \es x e^{-x}~, \nnb \\
F_5 (x) \es e^{-x}~,
\eea

and
\bea
u_0 = \frac{M_2^2}{M_1^2+M_2^2}~,~~~~~~M^2=\frac{M_1^2 M_2^2}{M_1^2+M_2^2}~,
\nnb
\eea
where $M_1^2$ and $M_2^2$ are the Borel parameters. Since the initial and
final baryons are the same,
we will set $M_1^2=M_2^2=2 M^2$, from which it follows that $u_0=1/2$. 

It follows from Eq. (\ref{sm1}) that the overlap amplitude
$\lambda_{\Lambda_Q}$ needs to be known in order to calculate the magnetic
moment. This amplitude is determined from baryon mass sum rules. 
For the mass sum rules of baryons we get (see also \cite{R19})
\bea
\label{sm2}
\lefteqn{
m_{\Lambda_Q} \lambda_{\Lambda_Q}^2 \, e^{-m_{\Lambda_Q}^2/M^2}
= \frac{m_Q}{32 \pi^4} 
(-13 + 2 b + 11 b^2) M^6 \Psi(3,0,m_Q^2/M^2) } \nnb \\ 
\ar \frac{\qq}{12 \pi^2} (1 + 4 b - 5 b^2) M^4 \Psi(1,-1,m_Q^2/M^2) \nnb \\
\ek \frac{m_Q}{36 M^2} m_0^2 \qq^2 \Big\{3 (5 + 2 b + 5 b^2) 
\Big[ F_4(m_Q^2/M^2)-F_5(m_Q^2/M^2)\Big] \\
\ar (-1+b)^2 F_5(m_Q^2/M^2) \Big\} \nnb \\ 
\ar \frac{\qq}{96 \pi^2} m_0^2 M^2 \left[(-1-4 b + 5 b^2) 
F_5(m_Q^2/M^2) + (-5 + 4 b + b^2) \Psi(1,0,m_Q^2/M^2) \right] \nnb \\ 
\ar \frac{m_Q}{6} \qq^2 (5 + 2 b + 5 b^2) F_5(m_Q^2/M^2)~, \nnb \\ \nnb \\
\label{sm3}
\lefteqn{
\lambda_{\Lambda_Q}^2 \, e^{-m_{\Lambda_Q}^2/M^2}
= \frac{3}{32 \pi^4} (5 + 2 b + 5 b^2) M^6
\Psi(3,-1,m_Q^2/M^2)} \nnb \\
\ek \frac{m_0^2}{72 M^2} \qq^2 \left[ (-26 + 4 b + 22 b^2) F_4(m_Q^2/M^2) + 
(-1+b)^2 F_5(m_Q^2/M^2) \right] \nnb \\ 
\ar \frac{m_Q}{12 \pi^2} \qq (1 + 4 b - 5 b^2) M^2 \Psi(2,0,m_Q^2/M^2) \\ 
\ar \frac{\qq^2}{18} (-13 + 2 b + 11 b^2) F_5(m_Q^2/M^2) \nnb \\ 
\ar \frac{m_Q}{96 \pi^2} m_0^2 \qq \left[ (-1  - 4 b + 5 b^2)
\Psi(1,0,m_Q^2/M^2) + 
6 (-1+b^2) \Psi(2,1,m_Q^2/M^2) \right]~. \nnb 
\eea
Eqs. (\ref{sm2}) and (\ref{sm3}) correspond to the structures proportional
to the unit operator and $\not\!p$, respectively.  
Subtraction of the continuum contribution in Eqs. (\ref{sm1}), (\ref{sm2})
and (\ref{sm3}) can be performed by making the following substitution
\bea
M^{2n} \psi(\alpha,\beta,m_Q^2/M^2) \rightarrow  
\frac{1}{\Gamma(\alpha)} \frac{1}{\Gamma(n)} \int_{m_Q^2}^{s_0} ds e^{-s/M^2} 
\int_1^{s/m_Q^2} dt (s - t m_Q^2)^{n-1} t^{\beta - \alpha -1} 
(t-1)^{\alpha - 1}~,
\eea
for $\alpha > 0$ and $n > 0$.
\section{Numerical analysis}

This section is devoted to the numerical analysis of the sum rules for the
magnetic moment of $\Lambda_Q$ baryon. It follows from Eq. (\ref{sm1}) that
the main input parameter of the LCQSR is the photon wave function. It was
shown in \cite{R13,R14} that the leading twist--2 photon wave functions
receive minor corrections from the higher conformal spin, so that to a good
approximation they can be described by their asymptotic forms. In further
numerical calculations we will use the following forms of the photon wave
functions \cite{R13,R15}:
\bea
\phi(u) \es 6 u (1-u)~,~~~~~\psi(u) = 1~, \nnb \\
g_1(u) \es - \frac{1}{8}(1-u)(3-u)~,~~~~~g_2(u) = -\frac{1}{4} (1-u)^2~.\nnb
\eea
The values of the other input parameters which we need in 
the numerical calculations are:
$f=0.028~GeV^2$, $\chi=-4.4~GeV^{-2}$ \cite{R20} (in  \cite{R21} this
quantity is estimated to be $\chi=-3.3~GeV^{-2}$), 
$\qq(1 ~ GeV)=-(0.243)^3~GeV^3$, $m_0^2=(0.8\pm0.2)~GeV^2$ \cite{R22},
$m_c=1.3~GeV$ and $m_b=4.8~GeV$.

In view of the fact that the 
magnetic moment is a physical quantity it must be independent of the
auxiliary parameters $t$, Borel mass $M^2$, as well as the continuum threshold
$s_0$. Therefore calculation of the magnetic moment requires simply finding a
region where $\mu_{\Lambda_Q}$ is practically independent of the 
above--mentioned parameters.

Our strategy in trying to resolve this problem consists of three steps. In the first
step, in order to find the working region, where $\mu_{\Lambda_Q}$ is
supposed to be independent of the Borel parameter $M^2$,  we study the dependence 
of the magnetic moment $\mu_{\Lambda_Q}$ on $M^2$ at several predetermined values of 
the threshold $s_0$ and at three different values of $t$. Along these lines,
in Fig. (1) we present the dependence of $\mu_{\Lambda_c}$ on $M^2$ at
$s_0=10~GeV^2$ and $s_0=15~GeV^2$, for different values of $t$. 
Similarly, Fig. (2) depicts the
dependence of $\mu_{\Lambda_b}$ on $M^2$ at two different values 
of the threshold, 
$s_0=40~GeV^2$ and $s_0=45~GeV^2$.
From Fig. (1) we observe that the working region of $M^2$ which is
consistent with our requirements lies in the interval 
$3~GeV^2 \le M^2 \le 5~GeV^2$ for the $\Lambda_c$ case and 
$15~GeV^2 \le M^2 \le 25~GeV^2$ for the $\Lambda_b$ case. Obviously, from
both these figures we observe that $\mu_{\Lambda_Q}$ is reasonably 
insensitive to the different choices of the continuum threshold, while it 
seems to be sensitive to the parameter $t$.

The next step, of course, is to explore the physical region for the
parameter $t$. For this purpose we have used the mass sum rules given in
Eqs. (\ref{sm2}) and (\ref{sm3}), both of which are required to be positive.
Following this line of reasoning, we present in Figs. (3) and (4) the
dependence of the mass sum rule (\ref{sm2}) on $\cos \theta$, where  
$\theta$ is determined from the relation $\tan \theta=t$, for $\Lambda_c$
and $\Lambda_b$, respectively. These figures depict that the relevant
physical regions for the parameter $t$ are given by 
$-0.78 \le \cos\theta \le 0.7$ for
$\Lambda_c$ and  $-0.8 \le \cos\theta \le 0.7$ for $\Lambda_b$, respectively.
The analysis of (\ref{sm3}), clearly, leads to the conclusion
that this sum rule is positive for arbitrary values of the parameter $t$. 
As a result of these observations, the physical region of $t$ which
guarantees the positiveness of the sum rules (\ref{sm2}) and (\ref{sm3})
separately, is confined to the interval $-0.78 \le \cos\theta \le 0.7$. Having this
restriction on $t$, our final attempt is to determine the value of the
magnetic moment $\mu_{\Lambda_Q}$. For this purpose, we must find a region
of $t$ where $\mu_{\Lambda_Q}$ is independent of this parameter.

In Figs. (5), we present the dependence of the magnetic moment of 
heavy baryon $\Lambda_c$ on $\cos \theta$ at the fixed
value of the Borel parameter $M^2=4~GeV^2$ for two different choices of the
continuum threshold $s_0=3~GeV^2$ and $s_0=4~GeV^2$. Similarly, depicted in Fig. (6)
is the dependence of the other heavy baryon $\Lambda_b$ on $\cos \theta$ at
$M^2=20~GeV^2$ and $s_0=40~GeV^2;~s_0=45~GeV^2$. We observe from these
figures that the magnetic moment is quite stable in the region 
$-0.25 \le \cos \theta \le +0.5$, and practically seems to be independent
of $\cos \theta,~t$ and the continuum threshold $s_0$. As a result of all
these considerations we obtain for the magnetic moment
\bea
\label{mlclb1}
\mu_{\Lambda_c} \es \ga 0.40 \pm 0.05 \dr \mu_N ~, \nnb \\
\mu_{\Lambda_b} \es \ga -0.18 \pm 0.05 \dr \mu_N ~,
\eea
where $\mu_N$ is the nucleon magneton.

Finally we present a comparison of our result on $\mu_{\Lambda_Q}$ with the
existing theoretical calculations in literature. For the magnetic moment 
$\mu_{\Lambda_c}$ the traditional QCD sum rules predicts 
$\mu_{\Lambda_c}= \ga 0.15 \pm 0.05 \dr \mu_N$ \cite{R5}. Our result on 
$\mu_{\Lambda_c}$ is close to the non--relativistic quark model 
prediction \cite{R23,R24}, but there is substantial difference with the
results predicted in \cite{R5}. In our opinion, this discrepancy can be 
attributed to the fact that the results presented in 
\cite{R5} were calculated for the choice $t=-1$, which is unphysical in our
case.
Magnetic moments of triplet and sextet heavy baryons have been calculated
using heavy hadron chiral perturbation theory (HHChPT) in
\cite{R25,R26,R27}, and $\Lambda_c$ and $\Lambda_b$ baryons that we are
interested belong to the triplet. It was shown in these works that in
HHChPT the leading term in magnetic moments is proportional to $e_Q/m_c$,
where $Q$ is the heavy quark. The corrections to the leading term
appear at the order of ${\cal O}(1/m_Q \Lambda_\chi^2)$, where
$\Lambda_\chi \sim 1~GeV$, and four arbitrary constants are required.
If we set all four arbitrary constants to zero, for the magnetic moments of
$\Lambda_c$ and $\Lambda_b$ baryons HHChPT predicts 
\bea
\label{mlclb2}
\mu_{\Lambda_c} \se 0.47~\mu_N~, \nnb \\
\mu_{\Lambda_b} \se -0.23~\mu_N~.
\eea
From Eqs. (\ref{mlclb1}) and (\ref{mlclb2}) we conclude that both approaches
lead to close values for the magnetic moments of the heavy $\Lambda_c$ and
$\Lambda_b$ baryons. The small difference between the predictions on the
magnetic moments of heavy baryons of the two approaches is due to the
subleading terms which are neglected in deriving Eq. (\ref{mlclb2}).

A measurement of the magnetic moments of heavy baryons represents an
experimental challenge. Few groups are contemplating the possibility of
performing magnetic moments in the near future (BTeV and SELEX)
\cite{R28}.

\newpage

\section*{Figure captions}
{\bf Fig. (1)} The dependence of the magnetic moment 
$\mu_{\Lambda_c}$ on
$M^2$ at two different values of the
continuum threshold $s_0=10~GeV^2$ and $s_0=15~GeV^2$, for several fixed 
values of the parameter $t$. Here in this figure and in all following figures
the magnetic moments of $\Lambda_c$ and $\Lambda_b$ baryons are given in
units of the nucleon magneton $\mu_N$. \\ \\
{\bf Fig. (2)} The dependence of the magnetic moment
$\mu_{\Lambda_b}$ on
$M^2$ at two different values of the   
continuum threshold $s_0=40~GeV^2$ and $s_0=45~GeV^2$, for several fixed
values of the parameter $t$.\\ \\
{\bf Fig. (3)} The dependence of the mass sum rule 
$m_{\Lambda_c}$ on 
$\cos \theta$, at two different values of the   
continuum threshold $s_0=10~GeV^2$ and $s_0=15~GeV^2$.\\ \\
{\bf Fig. (4)} The dependence of the mass sum rule 
$m_{\Lambda_b}$ on 
$\cos \theta$, at two different values of the   
continuum threshold $s_0=40~GeV^2$ and $s_0=45~GeV^2$.\\ \\
{\bf Fig. (5)} The dependence of the magnetic moment 
$\mu_{\Lambda_c}$ on $\cos \theta$, at $M^2=4~GeV^2$ and
at two different values of the
continuum threshold $s_0=3~GeV^2$ and $s_0=4~GeV^2$.\\ \\
{\bf Fig. (6)} The dependence of the magnetic moment
$\mu_{\Lambda_b}$ on
$\cos \theta$, at $M^2=20~GeV^2$ and 
at two different values of the
continuum threshold $s_0=40~GeV^2$ and $s_0=45~GeV^2$.
\newpage

\begin{figure}
\vskip 1.5 cm
    \includegraphics{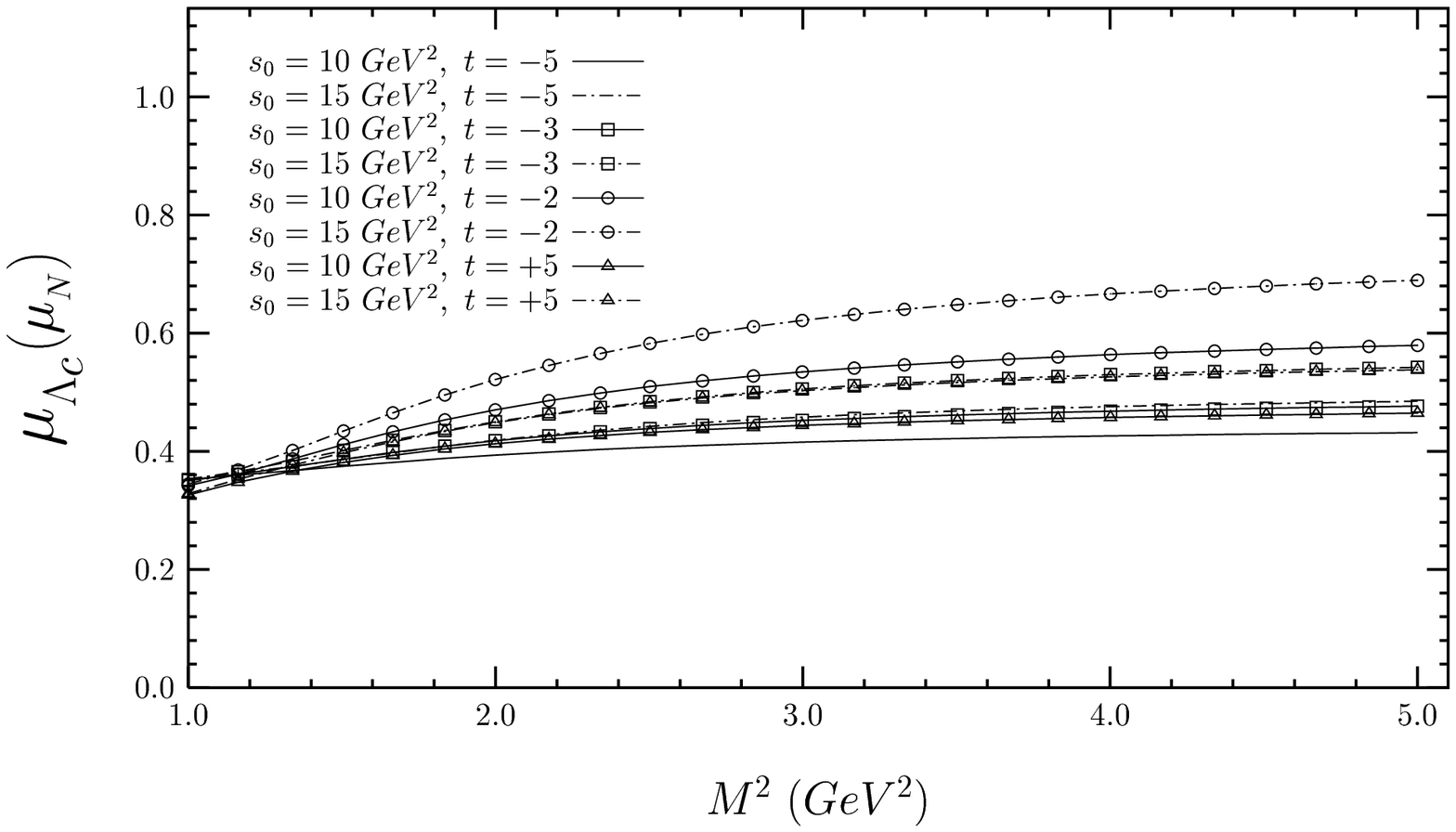}
\vskip 7.8cm
\caption{}
\end{figure}

\begin{figure}  
\vskip 2.5 cm
    \includegraphics{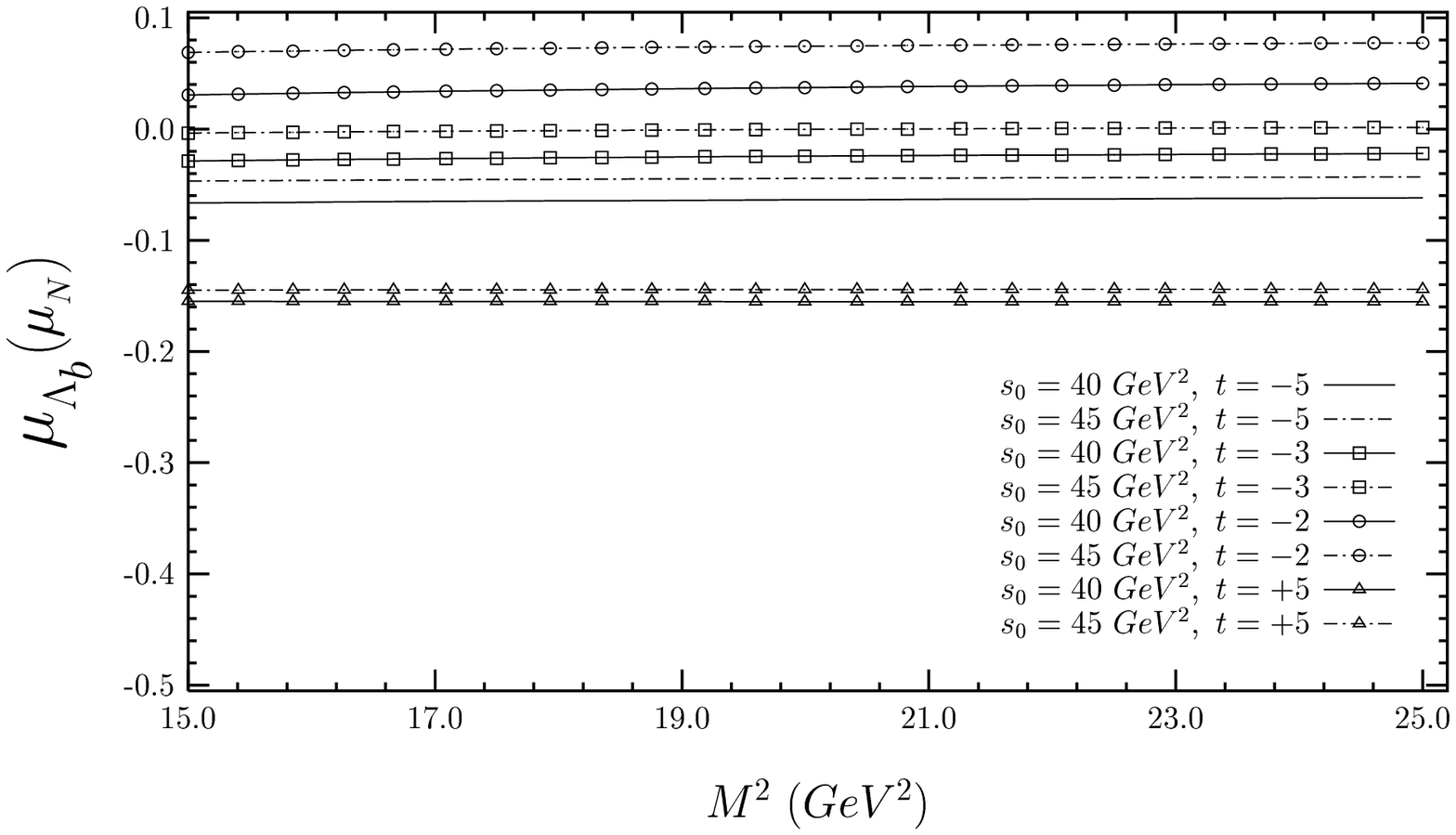}
\vskip 7.8 cm   
\caption{}
\end{figure}

\begin{figure}
\vskip 1.5 cm
    \includegraphics{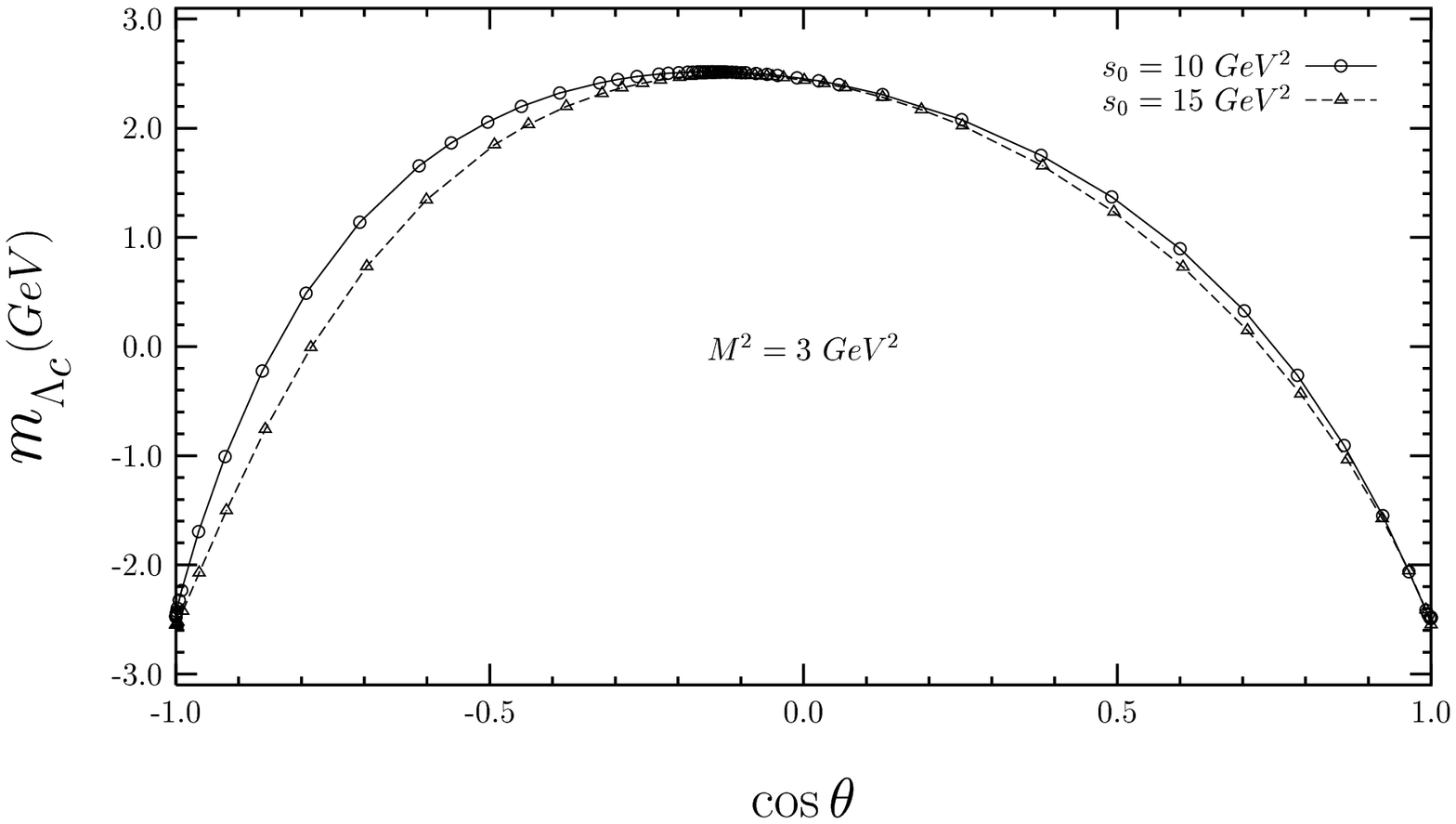}
\vskip 7.8cm
\caption{}
\end{figure}

\begin{figure}  
\vskip 2.5 cm
    \includegraphics{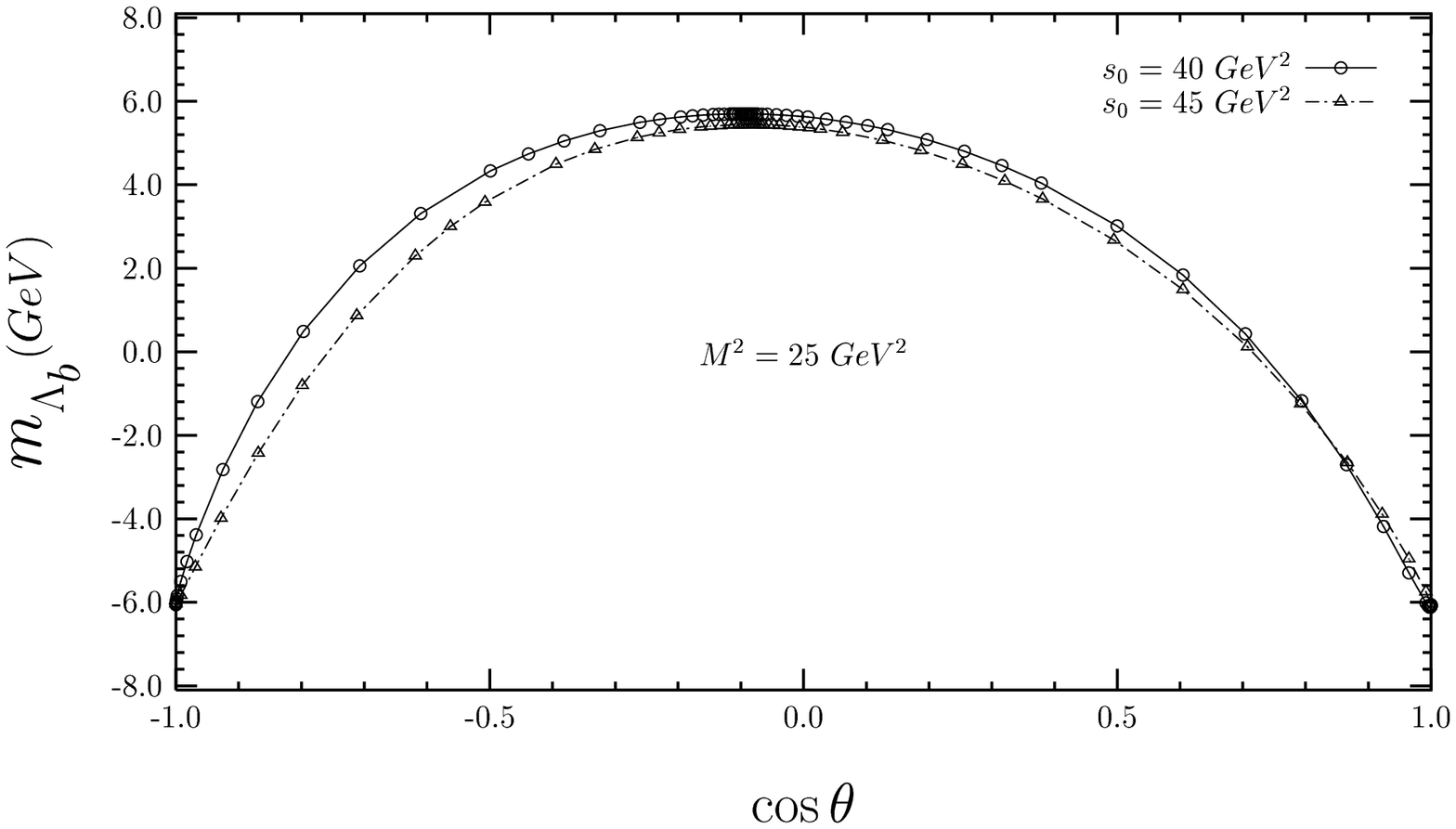}
\vskip 7.8 cm   
\caption{}
\end{figure}

\begin{figure}
\vskip 1.5 cm
    \includegraphics{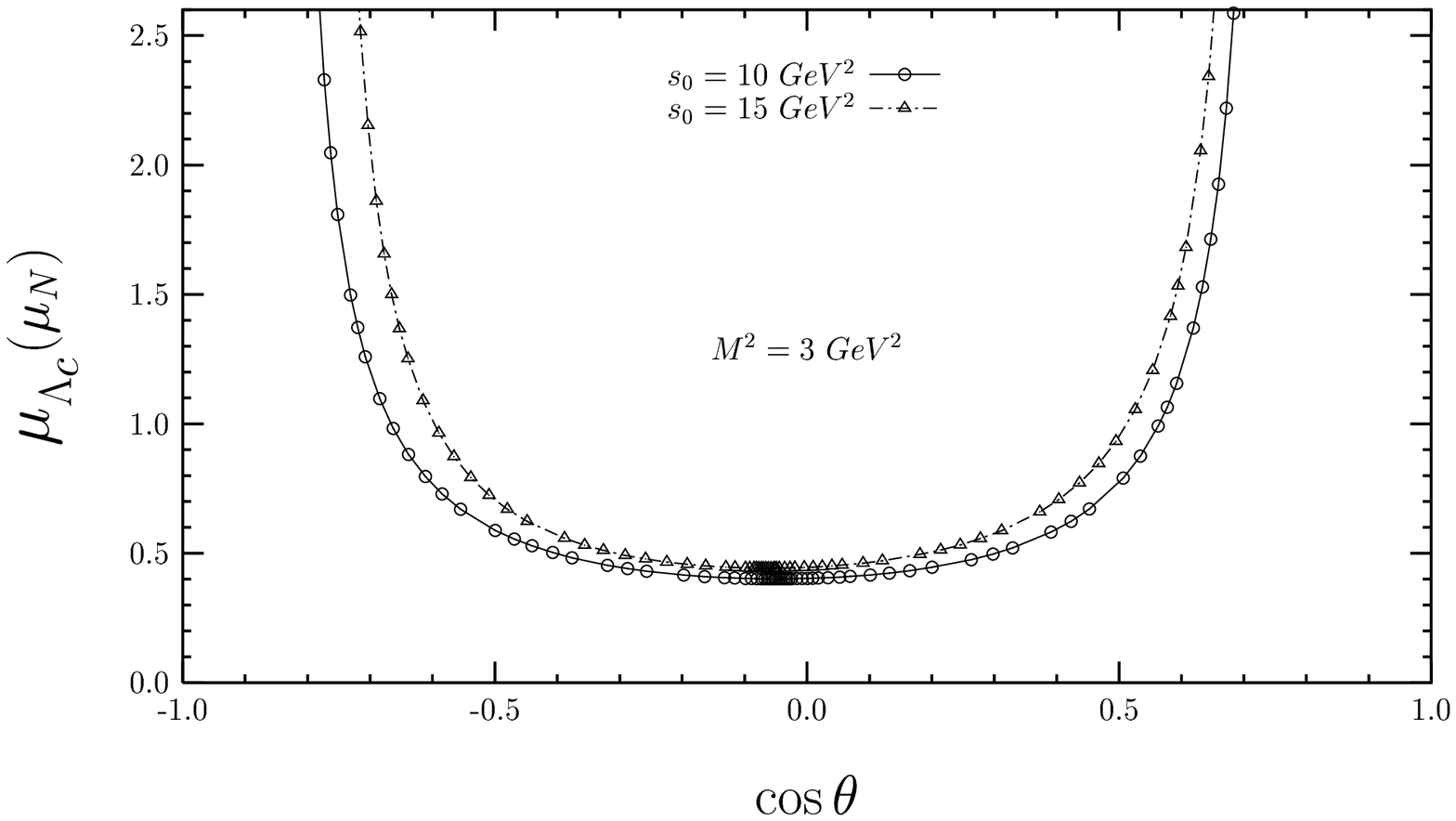}
\vskip 7.8cm
\caption{}
\end{figure}

\begin{figure}  
\vskip 2.5 cm
    \includegraphics{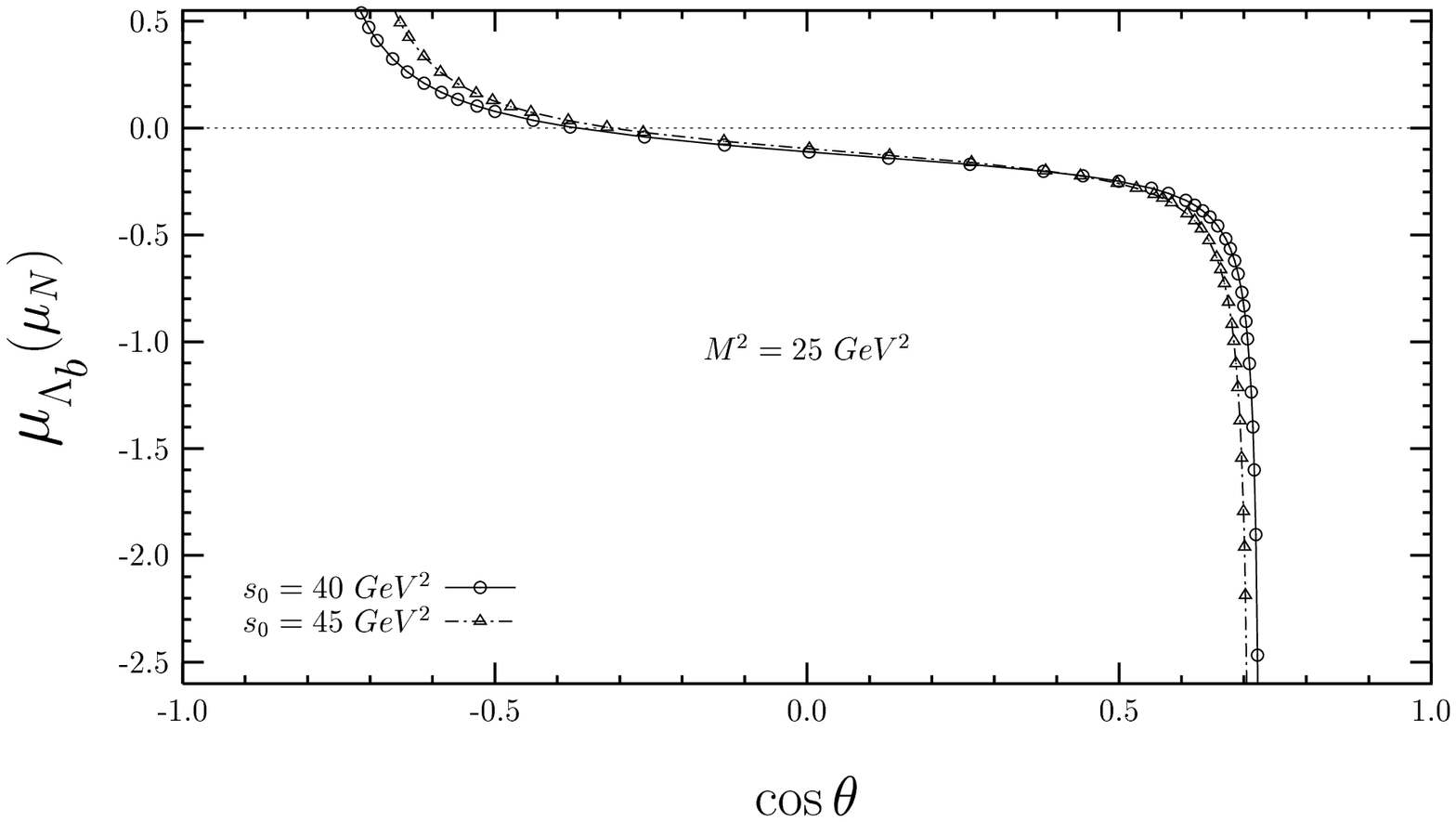}
\vskip 7.8 cm   
\caption{}
\end{figure}

\end{document}